\documentclass[aps,prb,twocolumn,superscriptaddress,showpacs]{revtex4}
\usepackage{graphicx}
\newcommand{\beq}{\begin{equation}}
\newcommand{\eeq}{\end{equation}}
\begin{document}


\title{Aharonov-Bohm effect for pedestrian}

\author{J. Planelles}
\email{planelle@exp.uji.es}
\affiliation{Departament de Ci\`encies Experimentals, Universitat Jaume I, Box 224, E-12080 Castell\'o, Spain.}

\author{J. I. Climente}
\affiliation{INFM-National Research Center on nano-Structures and bio-Systems at Surfaces (S$^3$),
Universit\`a degli Studi di Modena e Reggio Emilia, Via Campi 213/A, 41100 Modena, Italy.}
\affiliation{Departament de Ci\`encies Experimentals, Universitat Jaume I, Box 224, E-12080 Castell\'o, Spain.}

\author{J. L. Movilla}
\affiliation{Departament de Ci\`encies Experimentals, Universitat Jaume I, Box 224, E-12080 Castell\'o, Spain.}

\begin{abstract}
When a magnetic field pierces a multiple-connected quantum system, the corresponding wavefunction is altered although no net Lorentz force acts upon its carriers. This is the so called Aharonov-Bohm effect. The most simple multiply-connected quantum system is a quantum ring QR. Nowadays it is possible to obtain QRs in the nanoscopic range providing spectroscopic data vs. and applied external magnetic field. We describe here the most significant quantum effects induced by the magnetic field in a QR by means of simple quantum mechanical models.\\

\end{abstract}
\maketitle
\section{Introduction}
In their celebrated 1959 paper\cite{AB} Aharonov and Bohm pointed out that while the fundamental equations of motion in classical mechanics can always be expressed in terms of field alone, in quantum mechanics the canonical formalism is necessary, and as a result, the potentials cannot be eliminated from the basic equations. They proposed several experiments and showed that an electron can be influenced by the potentials even if no fields acts upon it. More precisely, in a field-free multiply-connected region of space, the physical properties of a system depend on the potentials through the gauge-invariant quantity $\oint A dl$, where $A$ represents the potential vector.\\

The most simple multiple-connected quantum system is a quantum ring, QR. Over the last two decades there has been an impressive experimental development towards smaller QRs. Early experiments reported observations of Aharonov-Bohm (AB) oscillations and persistent currents in mesoscopic metallic and semiconductor rings\cite{webb,levy,webb2,fuhere} where scattering still influences the phase coherent transport and a large number of electrons are present. More recently, Lorke et al. \cite{emperador,lorke} obtained self-assembled InAs semiconductor QRs in the nanoscopic range, each of which charged with one\cite{emperador} and two electrons\cite{lorke}, providing spectroscopic data in the scatter-free limit as a function of an external magnetic field. Simple two-dimensional effective mass models with parabolic-like spatial confinement \cite{emperador,lorke,hu,puente} yield reasonable agreement with most of experimental data, although truly 3D models are required to properly account for the vertical dimension and the Coulomb interaction\cite{plane1,naxo1,naxo2,naxo3,plane3} which is systematically overestimated by 2D models, as they miss vertical motion.\\

There is by now a vast literature both experimental and theoretical on QRs, including simple 1D models which can grasp basic behaviors of these multiply-connected systems \cite{viefers}. In the present paper we revisit at an elementary level the most significant quantum effects produced by a magnetic field on a quantum ring. 

\section{Aharonov-Bohm effect}

The Hamiltonian of a charged particle in a magnetic field reads,
\beq
\label{abe1}
\hat{\cal H}=\frac{(\hat{p}-eA)^2}{2 m_e} +V
\eeq
\noindent where $\hat p$ is the canonical moment, $A$ the potential vector, $e$ the particle charge and $V$ the spatial confining potential. If the magnetic field is axial and constant, $\vec B= B_0 \vec k$, we may choose the potential vector $\vec A= (-\frac{1}{2} y \, B_0, \frac{1}{2} x \, B_0, 0)$ so that the Hamiltonian eq. \ref{abe1} turns into:

\begin{eqnarray}
\label{abe2}
\hat{\cal H}&=&-\frac{\hbar^2}{2 m_e}\nabla^2 - \frac{e B}{2 m_e}\hat L_z+ \frac{e^2 B^2}{8 m_e} \rho^2 +V \nonumber \\
&=& \frac{\hat p_z^2}{2 m_e}+ \hat{\cal H}_{HO}^{2D} - \frac{e B}{2 m_e}\hat L_z +V
\end{eqnarray}

\noindent where $\hat{\cal H}_{HO}^{2D}$ is the 2D harmonic oscillator Hamiltonian.\\

If the vertical confinement is severe so that we can approximately separate variables and only consider the vertical ground state, and additionally, the in-plane confinement is zero or parabolic, the eigenvalues (Landau levels) grow linearly\footnote{In actual 3D confinements it may grow quadratically\cite{plane3}.} with the magnetic field and never intersect.\\

Now, if the particle is spatially confined in a hollow cylinder and we apply an axial magnetic field inside the inner radius $a$ only, i.e., $B=B_0$ if $0<\rho<a$ and  $B=0$ otherwise, we may choose the following potential vector:
\beq
\label{abe3}
\vec{A}=A_{\phi} \vec{u}_{\phi}=\left\{\begin{array}{ll}
\frac{1}{2}\,B \, \rho\, \vec{u}_{\phi} & 0<\rho<a\\
\frac{B a^2}{2 \rho} \, \vec{u}_{\phi} & a<\rho<\infty\\ \end{array} \right.
\eeq
\noindent which is continuous at $\rho=a$, where $B$ has a step-like discontinuity\footnote{Note that although we may choose another potential vector yielding the same magnetic field, no gauge will allow us to select $A=0$ in all the region where the system is located because the gauge-invariant flux $\Phi=\int B dS=\oint A dl$ is not zero.}. The selected potential vector fulfills the Coulomb gauge, $\nabla A=0$. Then,  $\hat p$ and $A$ commute and the Hamiltonian eq. \ref{abe1} describing our system (which is located in the interval $a<\rho<\infty$) becomes:

\beq
\label{abe4}
\hat{\cal H}=-\frac{\hbar^2}{2 m_e}\nabla^2 +\frac{i \hbar e}{m_e} \frac{B a^2}{2\rho^2}\frac{\partial}{\partial \phi}
+ \frac{e^2 B^2 a^4}{8 m_e \rho^2}  +V
\eeq
This Hamiltonian can also be obtained by formally replacing,
\beq
\label{abe5}
\frac{\partial}{\partial \phi} \to \frac{\partial}{\partial \phi} -  \frac{i \, e B \, a^2}{2\, \hbar}=\frac{\partial}{\partial \phi} + i \frac{\Phi}{\Phi_0}
\eeq
\noindent in the zero field Hamiltonian. In the above eq. \ref{abe5}, $\Phi=\pi a^2 B$ is the magnetic flux and $\Phi_0=2 \pi \hbar/|e|$ the flux unit.\\

Since the system has axial symmetry, the wave function can be written $\Psi(\rho,z) e^{i m \phi}$. Then, the presence of magnetic field inside the inner cylinder radius is accounted by the replacement $ m \to m + \frac{\Phi}{\Phi_0}$ in the differential equation on $(\rho,z)$ which yields eigenvalues. Therefore, if the magnetic field fulfills $\Phi=n \Phi_0$, $n=1,2,3 \dots$ we will get the same energies as those at $B=0$.\\

\enlargethispage{\baselineskip}
\noindent In the simple case of an electron in a 1D QR, the Hamiltonian is (a.u.):
\beq
\label{abe7}
\hat{\cal H}=-\frac{1}{2 m_e^* R^2} \left(\frac{\partial}{\partial \phi}+i \frac{\Phi}{\Phi_0}\right)^2
\eeq
\noindent and the energies,
\beq
\label{abe8}
E_m=\frac{1}{2 m_e^* R^2}(m+F)^2,
\eeq
\noindent where $m_e^*$ is the electron effective mass, $F=\frac{\Phi}{\Phi_0}$, and $m=0 \pm 1 \pm2 \dots$ The $E_m$ vs. $F$ plotting shows periodic energy intersections and changes in the $m$ symmetry of the ground state (AB effect).\\

\section{Fractional Aharonov-Bohm effect}

The Hamiltonian of two electrons in a 1D QR pierced by a magnetic field read in atomic units (a.u.):
\begin{eqnarray}
\label{abf1}
\hat{\cal H}(1,2) &=& -\frac{1}{2 R^2} \left(\frac{\partial}{\partial \phi_1}+i F \right)^2 \nonumber \\
 & & -\frac{1}{2 R^2} \left(\frac{\partial}{\partial \phi_2}+i F \right)^2+ \frac{1}{r_{1 2}}
\end{eqnarray}
\noindent with $ r_{1 2} = 2 \, R |\sin \frac{\phi_2-\phi_1}{2}|$, and where we assume, without loss of generality, that the electron effective mass $m_e^*=1$.\\

Disregarding the Coulomb interaction by the time being, the energy eigenvalues are:
\beq
\label{abf2}
E(m_1,m_2)=\frac{1}{2 R^2} \left[ (m_1+F)^2+(m_2+F)^2 \right]
\eeq
\noindent which show periodic changes of the ground state at the same values of flux as in the one-electron case. The eigenfunctions are either singlets ($S$) or triplets ($T$). The corresponding unnormalized spacial parts are
\beq
\label{abf3}
|m_1,m_2;S/T \rangle =e^{i m_1\phi_1} e^{i m_2\phi_2}\pm e^{i m_1\phi_2} e^{i m_2\phi_1}.
\eeq
Note that $|m,m;T \rangle$ does not exist (is zero). Then, eq. \ref{abf2} evidences that independently of the magnetic flux, the ground state is always singlet. At $F=\frac{1}{2},\frac{3}{2},\frac{5}{2} \dots$ the ground state is degenerate (three singlets and a triplet).\\

Prior to include the Coulomb term it is worthwhile to solve this problem again in a new set of coordinates:
\beq
\label{abf4}
s=\frac{1}{2}(\phi_1+\phi_2) \;\;\;\;\;\;\;\;\;\;\;\;\;\;\;\; r=\frac{1}{2}(\phi_1-\phi_2).
\eeq	
The spatial part of the eigenfunctions are now,
\begin{eqnarray}
\label{abf6}
|M,m;S\rangle &=& e^{i M s} \cos m r, \nonumber \\
|M,m;T\rangle &=& e^{i M s} \sin m r,
\end{eqnarray}
\noindent where $M=m_1+m_2$, $m=m_1-m_2$ and therefore, no triplets can be associated with $m=0$.
Since $M+m=2 m_1$ and $M-m=2 m_2$, we conclude that $M$ and $m$ must have same parity. In terms of $M$ and $m$ the energy reads:
\beq
\label{abf8}
E(M,m)=\frac{1}{4 R^2}\left[(M+2\,F)^2+m^2 \right],
\eeq
\noindent showing that the ground state energy at integer values of magnetic flux is zero and corresponds to singlet states.\\

\begin{figure}[h]
\begin{center}
\includegraphics[width=0.45\textwidth]{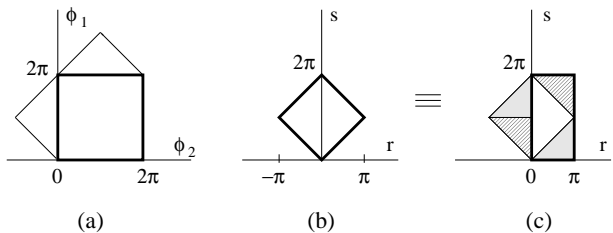}
\caption{\footnotesize Mapping between ($\phi_1$,$\phi_2$) and ($r$,$s$) domains.}\label{fig2}
\end{center}
\end{figure}

In terms of these new coordinates the Hamiltonian (disregarding the coulomb term) reads,
\beq
\label{abf9}
\hat{\cal H}(1,2)=-\frac{1}{4 R^2} \left(\frac{\partial}{\partial s}+2\,i F \right)^2 -\frac{1}{4 R^2} \frac{\partial^2}{\partial r^2}= \hat{\cal H}_s+\hat{\cal H}_r
\eeq
The $\hat{\cal H}_s$ eigenvectors, $e^{i M s}$, and eigenvalues,
\beq
\label{abf10}
E(M)=\frac{1}{4 R^2}(M+2\,F)^2,
\eeq
\noindent describe the dynamic of the center of mass (CM). The allowed values for $M$ will be fixed by the boundary conditions (BCs). Note that the electron permutation does not change the coordinate $s$. Therefore, $e^{i M s}$ is symmetric, and as a consequence, we must select a relative motion eigenfunctions ($\hat{\cal H}_r$ eigenfunctions) either symmetric (for singlets) or antisymmetric (for triplets). From the degenerate set $e^{\pm i m r}$ we choose $\cos m r$ and $\sin m r$ for singlets/triplets, respectively. The values that $m$ can reach will also be fixed by the BCs. However, as  shown in Figure \ref{fig2}, the domain of $r$ and $s$ are not independent so that BCs must be imposed upon the full spatial wave function. \\

The periodic BCs  $\phi_1\equiv \phi_1+ 2 \pi$, $\phi_2\equiv \phi_2+ 2 \pi$ yield $(s,r) \equiv (s+\pi, r+ \pi)$ and $(s,r) \equiv (s+\pi, r- \pi)$, i.e.,
\beq
\label{abf11}
e^{i\,M\,s} \left\{ \begin{array}{c} \sin m r \\ \cos m r \end{array} \right.  \equiv e^{i\,M\,s} e^{i\,M\,\pi} \left\{\begin{array}{c} \sin m (r \pm \pi) \\ \cos m (r \pm \pi)  \end{array} \right.
\eeq
Then, $m$ must be integer. Since $m \in Z$, then, eq. \ref{abf11} can be rewritten,
\beq
\label{abf12}
1=e^{i\,M\,\pi} \cos m  \pi, 
\eeq
\noindent which shows that $M$ must also be integer and that $M,m$ must have same parity.\\

Let us next include the coulomb term. It does not modify $\hat{\cal H}_s$, while $\hat{\cal H}_r$ becomes:
\beq
\label{abf13}
\hat{\cal H}_r=-\frac{1}{4 R^2 } \frac{\partial^2}{\partial r^2}+\frac{1}{2\,R |\sin r|} 
\eeq
No analytical solutions can be obtained (see however ref. \cite{zhu,truong}). The potential term $1/ \sin r$ defines a natural domain $0<r<\pi$ so that $\Psi_n(0)=\Psi_n(\pi)=0$ are the implicit (natural) BCs. Within this domain the eigenfunctions which are non-degenerate, show the correct nodal sequence. However, we stated above that $-\pi<r<\pi$ and, additionally, $r$ and $s$ domains are not independent. We may use the periodicity of the problem to select the domains $0<s<2 \pi$, $0<r<\pi$ (see Fig \ref{fig2}c).\\

We study next the Pauli's principle restrictions in the presence of Coulomb interactions. As before, $e^{iMs}$ is invariant under the particles permutation operator ${\cal P}_{12}$. On the other hand, as ${\cal P}_{12} r= -r$, it is followed that
${\cal P}_{12} \Psi(r)= \Psi(-r)$. As it was stated before, the periodicity of our problem allows to stablish the equivalence $(s,r) \equiv (s+\pi,r+\pi)$ and therefore,
\beq
\label{abf15}
e^{i\,M\,s} \Psi_n(-r)=e^{i\,M\,s} e^{i\,M\,\pi} \Psi_n(\pi-r) 
\eeq
\noindent i.e.,
\beq
\label{abf16}
\hat {\cal P}_{12} \Psi_n(r)=(-1)^M \Psi_n(\pi-r).
\eeq
The symmetry of $\Psi_n(r)$ with respect to $r=\pi/2$ and the nodal sequence of eigenvectors allows us to write $\Psi_n(\pi-r)=(-1)^n \Psi_n(r)$. Then eq. \ref{abf16} yields
\beq
\label{abf17}
\hat {\cal P}_{12} \Psi_n(r)=(-1)^{(M+n)} \Psi_n(r).
\eeq
If $M+n$ is even, the spatial function should be then symmetric (and then, its spin partner antisymmetric, i.e. singlet). On the contrary, $M+n$ odd corresponds to triplets.\\

In absence of magnetic flux the lowest CM state $|M=0\rangle$ and the lowest relative motion state $|n=0\rangle$ combine into the ground state (singlet). As the magnetic flux reaches $F=1/2$, then the lowest CM state is $|M=-1\rangle$ (see eq. \ref{abf10}). It combines with $|n=0\rangle$, which is flux independent, yielding a triplet ground state with the same energy as the singlet ground state at $F=0$. As the magnetic flux increases a new singlet becomes the ground state, then a triplet, etc. We see that Coulomb interactions halves the periodicity of the AB effect (fractional AB effect\cite{niemela}).\\

The coulomb interaction in a 1D system is actually unrealistically large. A very simple model accounting for the 3D character of a real QR may be represented by the Hamiltonian
\beq
\label{abf18}
{\cal H}_{\xi} =-\frac{1}{4 R^2} \frac{\partial^2}{\partial r^2}+\frac{1}{\xi+2\, R\,|\sin r|}
\eeq
\noindent where the parameter $\xi$ incorporates somehow the average of the Coulomb potentials over the coordinates $\rho$ and $z$. Note that if we set $\xi=\infty$ then ${\cal H}_{\xi}$ corresponds to an independent particles model, while the limit $\xi=0$  corresponds to interacting electrons in a 1D QR. The numerical integration of ${\cal H}_{\xi}$ assuming reasonable $\xi$ values shows that although triplets remain as ground states at fractional flux values, the energies are larger and the triplet windows shorter than those of singlets.

\section{Antiperiodic Boundary Conditions and the Aharonov-Bohm effect}

The M\"obius strip problem is of high theoretical interest since classically the AB periodicity is related to interference between trajectories. In a M\"obius strip the electron encircles the system twice before returning to its initial position. Then, we may expect differences between persistent currents in a M\"obius strip and a QR. Since a M\"obius strip cannot actually be pressed into a 1D structure, in order to isolate the effects, we will devote this section to study one and two electrons in a 1D QR with antiperiodic BCs (AQR). Let us consider first only one electron. The Hamiltonian is the same as in section 1 (eq. \ref{abe7}) and the antiperiodic BCs, $\Psi_m(\phi+2 \pi)=-\Psi_m(\phi)$ yield  $m=\pm 1/2, \pm 3/2, \pm 5/2,\dots$ Then, the $E_m$ vs. $F$ plotting is identical to that of one electron in a QR except for a shift of $1/2$ unit of flux.\\

In terms of the CM and relative motion coordinates $s$ and $r$, eq. \ref{abf4}, the wave functions of two non-interacting electrons are given by the same eq. \ref{abf6} as the QR, and again $M,m \in Z$ (because $M=m_1+m_2$ and $m=m_1-m_2$). However $M$ and $m$ must have opposite parity\footnote{We may write $m_1=(2 p +1)/2$, $m_2=(2 q +1)/2$ with $p,q \in Z$. Then, $M=p+q+1$ and  $m=p-q$. Since $p+q$ and  $p-q$ have same parity, then $M$ and $m$ must have it opposite. The same result can be obtained from the analogous of eq. \ref{abf11} for antiperiodic BCs yielding $1=-e^{i M \pi} \cos m \pi$ with $m \in Z$.}.\\

When the Coulomb term is included and, as in the previous section, we solve the problem in the domain $0<s<2 \pi$, $0<r<\pi$, we find analytical $e^{i M s}$ symmetric functions describing the CM motion and numerical $\Psi_n(r)$, $n=0,1,2,\dots$ functions for the relative motion. Again, ${\cal P}_{1 2} r = -r$ and then ${\cal P}_{1 2} \Psi_n(r) = \Psi_n(-r)$. However, eq. \ref{abf15} is replaced by:
\beq
\label{abc1}
e^{i\,M\,s} \Psi_n(-r)=-e^{i\,M\,s} e^{i\,M\,\pi} \Psi_n(\pi-r) 
\eeq
\noindent and then, eq.\ref{abf17} by
\beq
\label{abc2}
\hat {\cal P}_{12} \Psi_n(r)=(-1)^{(M+n+1)} \Psi_n(r),
\eeq
\noindent so that if $M+n$ is even/odd $|M,n\rangle$ will be triplet/singlet. Then, from eq. \ref{abf10}, we see that if $F=0$ the lowest $M=0$ will combine with $n=0$ yielding a triplet ground state. At $F=1/2$ it is $M=-1$ which combines with $n=0$ yielding a singlet state, etc. Therefore, we find out again the same picture as in QR except for a shift of half flux unit. The similarities between the 1D QR and AQR remain if we consider ${\cal H}_{\xi}$, eq. \ref{abf18}.

\section{Optic Aharonov-Bohm effect: excitons}

An exciton in a QR is a neutral entity. Then, it should not be sensitive to the applied magnetic flux.  However different masses of electrons and holes  yield observable effects in realistic 3D QR. Namely,  dark exciton in some windows of magnetic field\cite{naxo1}. This is the so called optical AB effect\cite{govorov}. Romer and Raikh\cite{romer} employed a short-range e-h attractive potential in a 1D QR and conclude that the AB effects will be present if electron and hole can tunnel in the opposite directions and meet each other on the opposite side of the ring. However it seems that actual Coulomb terms prevent the ground state oscillations in 1D QR \cite{hu2,song,bayer}.\\

The Hamiltonian of an electron and a hole in a 1D QR pierced by a magnetic field reads:
\begin{eqnarray}
\label{oab2}
\hat{\cal H}&=&-\frac{1}{2 m_e^* R^2} \left(\frac{\partial}{\partial \phi_e} + i F \right)^2-\frac{1}{2 m_h^* R^2} \left( \frac{\partial}{\partial \phi_h} -i \, F \right)^2 \nonumber \\
& & - \frac{1}{2 R |\sin \frac{\phi_e-\phi_h}{2}|}
\end{eqnarray}
\noindent where $m_e$, $m_h$ are the electron/hole effective masses, both considered positive in this model. If we disregard by the time being the Coulomb attraction, $\Psi(\phi_e,\phi_h)= e^{i\,M_e \, \phi_e}\, e^{i\,M_h \, \phi_h}$ is the eigenfunction  associated to the eigenvalue:
\begin{eqnarray}
\label{oab3}
\lambda &=& E-E_g \nonumber \\
&=& \frac{1}{2 m_e^* R^2} (M_e+F)^2 +\frac{1}{2 m_h^* R^2} (M_h-F)^2,
\end{eqnarray}
\noindent where $E_g$ is the electron-hole energy gap and $M_e$, $M_h$ $=0 \pm 1 \pm 2 \dots$ The $E$ vs. $F$ plot shows periodic changes of ground state $(M_e, M_h)=$ $(0,0)$,$(-1,1)$,$(-2,2) \dots$However, $M_L=M_e+M_h$ is always zero. Then, the selection rule $M_L=0$ is fulfilled and there are not dark windows for luminescence, i.e., no optic AB effect can be seen. If we take into account that electron and hole have different effective masses, we may think that in a real QR electron and hole will follow different orbits. A very simple model of a 2D QR where electron and hole follow circular orbits with radii $R_e \neq R_h$ pierced by a magnetic field (including the region where the system is located) has been recently proposed\cite{govorov}. This allows to have different flux inside the electron and hole orbits: $F_e=\pi R_e^2 B/\Phi_0$, $F_h=\pi R_h^2 B/\Phi_0$. As a result, eq. \ref{oab3} turns into
\beq
\label{oab4}
E=E_g+ \frac{1}{2 m_e^* R_e^2} (M_e+F_e)^2 +\frac{1}{2 m_h^* R_h^2} (M_h-F_h)^2,
\eeq
\noindent which allows states with total angular momentum $M_L=M_e+M_h \neq 0$ to become the ground state within some flux windows. As the selection rule $M_L=0$ dramatically reduces the emission intensity in these regions of magnetic flux (dark windows), the optic AB effect can now be observed.\\
It is worth to stress that while in the case of standard AB effect we loose the perfect periodicity as the magnetic field pierces the region where the system is located, it is not possible to observe the optic AB effect unless B pierces the system (so that a flux net between electron and hole orbits exists and a different phase factor in one and other particle occurs).

\begin{acknowledgments}
Financial support from MEC-DGI project CTQ2004-02315/BQU and UJI-Bancaixa project P1-B2002-01 are gratefully acknowledged. A MECD of Spain  FPU grant is also acknowledged (JLM). This work has been supported in part (JIC) by the EU under the TMR network ``Exciting''.
\end{acknowledgments}

\end{document}